\documentclass[aps,prd,twocolumn,superscriptaddress,nofootinbib]{revtex4}
\usepackage{graphicx,epsfig,color}
\usepackage{bm}
\begin{document}
\title{Tetraquark-adequate QCD sum rules for quark-exchange processes}
\author{Wolfgang Lucha}
%\email[]{wolfgang.lucha@oeaw.ac.at}
\affiliation{Institute for High Energy Physics, Austrian Academy of 
Sciences, Nikolsdorfergasse 18,\\ 
A-1050 Vienna, Austria}
\author{Dmitri Melikhov}
%\email[]{dmitri\_melikhov@gmx.de}
\affiliation{Institute for High Energy Physics, Austrian Academy of 
Sciences, Nikolsdorfergasse 18,\\ 
A-1050 Vienna, Austria}
\affiliation{D.~V.~Skobeltsyn Institute of Nuclear Physics,
M.~V.~Lomonosov Moscow State University,\\ 
119991, Moscow, Russia}
\affiliation{Faculty of Physics, University of Vienna, Boltzmanngasse
5, A-1090 Vienna, Austria}
\author{Hagop Sazdjian}
%\email[]{sazdjian@ipno.in2p3.fr}
\affiliation{Institut de Physique Nucl\'eaire, Universit\'e Paris-Sud,
CNRS-IN2P3,\\
Universit\'e Paris-Saclay, 91405 Orsay, France}

\date{\today}

\begin{abstract}
We consider quark-hadron duality relations and QCD sum rules 
for correlators involving exotic tetraquark currents, here specializing to the case 
of quark-exchange processes. We point out
the differences they exhibit with respect to the cases involving 
ordinary bilinear quark currents. Based on the observation that
only diagrams possessing at least four-quark singularities can
contribute to the formation of tetraquark states, we show that the
quark-hadron duality relations and the corresponding sum rules
split into two non-overlapping relations.  
The ultimate tetraquark-adequate QCD sum rule is concerned only
with one of these relations, in which the operator product expansion
starts with diagrams of order $O(\alpha_s^2)$.
\end{abstract}
\maketitle

\section{Motivation} \label{s1}
In recent years, experimental data provided increasing evidence
for near-threshold hadron resonances with a favorable interpretation
as tetraquark and pentaquark hadrons, i.e., hadrons with minimal parton
configurations consisting of four and five quarks, respectively
\cite{Esposito:2016noz,ali,Olsen:2017bmm,guo,nora2019}.
The experimental progress has been escorted by extensive theoretical
studies aimed at understanding the possible nature and structure
of such exotic hadrons in QCD.

Numerous works deal with the application of Shifman-Vainshtein-Zakharov
(SVZ) sum rules \cite{QCD_SR} to tetraquark and pentaquark states
\cite{QCD_SR_1,QCD_SR_2} (and references therein). The method of SVZ
sum rules (or QCD sum rules) makes use of dispersion representations to
calculate QCD Green functions, or correlators, in two different ways:
First, by applying the Wilson operator product expansion (OPE), which
gives the OPE (theoretical) side of a QCD sum rule.
Second, by calculating the same Green function by the insertion
of a complete set of hadronic intermediate states; this yields the
hadron (phenomenological) side of the QCD sum rule.
For Green functions of bilinear meson or trilinear baryon
interpolating currents, the hadron continuum is counterbalanced by
the contribution of perturbative QCD diagrams above an appropriate
effective threshold. Making use of this property, one relates
parameters of the ordinary hadrons to the low-energy region of
perturbative QCD diagrams starting at order $O(\alpha_s^0)$
(hereafter referred to as $O(1)$ diagrams, $\alpha_s$ being the
strong coupling constant) and supplemented by appropriate condensate
contributions \cite{QCD_SR}.

Previous applications of SVZ sum rules to exotic states
\cite{QCD_SR_1,QCD_SR_2} have followed the same route by
calculating the $O(1)$ QCD diagrams (and in some cases also radiative
corrections) and the corresponding power corrections,
and borrowing the same criteria for continuum subtraction as
prescribed for the ordinary hadrons \cite{QCD_SR}.
As a result, the tetraquark or pentaquark properties have been
related to $O(1)$ QCD diagrams.\footnote{
A caveat about subtleties in the application of the SVZ sum rules
to exotic states has been raised in \cite{5qsr}, where it was noticed
that, depending on the way one treats the contributions of
two-hadron states on the phenomenological side of the QCD sum rule,
one arrives at different assignments of the pentaquark quantum numbers.
The origin of this problem, however, was not clarified.}

However, the case of exotic multiquark currents has a fundamental
difference compared to the ordinary currents: namely, for correlators
of exotic currents, the OPE side as well as the hadron side of SVZ sum
rules may be split into two non-overlapping classes of contributions,
with respect to their singularity structure.
Moreover, diagrams of each of these two classes on the OPE side and on
the hadron side satisfy QCD sum rules of their own, i.e., the quark-hadron
duality relation for the exotic correlator leads to
two independent SVZ sum rules for each class of contributions. Exotic
states contribute to only one of these sum rules, which we name,
in the case of tetraquarks, $T$-adequate QCD sum rule, and do not
contribute to the other sum rule. 
The OPE side of the $T$-adequate sum rule contains contributions of
only those diagrams which may participate 
%are recognized as possibly taking part 
in the formation of the tetraquark state. These are called $T$-phile diagrams
and are selected according to their singularity structure in the
Feynman diagrams of the four-point function, $\Gamma_{4j}$, of
quark bilinear color-singlet currents.
$T$-phile diagrams should have a four-quark $s$-channel cut;
such diagrams emerge at order $O(\alpha_s^2)$ and higher
\cite{lms_prd100}. Obviously, in order to obtain meaningful estimates
of the parameters of exotic states, one needs to use the $T$-adequate
QCD sum rule.

In fact, in the context of large-$N_c$ QCD, it is well known
that some classes of Feynman diagrams are not related to exotic
hadrons \cite{coleman,weinberg,knecht,cohen,maiani,lms_prd,lms_epjc,maiani2,lms_pos,lms_prd2018}. 
However, only recently the consequences of this property for the
formulation of sum rules for exotic states has been worked out:
In \cite{lms_prd100}, we have focused on quark-hadron duality relations
for %exotic 
correlation functions of the tetraquark currents and
have explicitly demonstrated for specific ``direct'' Green functions,
corresponding to processes where the initial and final states have
the same quark-antiquark bilinear flavor structures, that the $O(1)$ and
the $O(\alpha_s)$ contributions on the QCD side precisely cancel
against the two-hadron continuum contributions on the hadronic side,
as soon as SVZ sum rules for correlators of ordinary currents are used.
The OPE side of the $T$-adequate QCD sum rule has been shown to start
with specific nonfactorizable diagrams of order $O(\alpha_s^2)$.

In this paper, we derive the $T$-adequate QCD sum rules for
``recombination'' diagrams, corresponding to quark-exchange processes,
and prove that also in this case the OPE side of the $T$-adequate QCD
sum rule starts with diagrams of order $O(\alpha_s^2)$.
For the sake of clarity, we discuss the case of exotic currents
composed of quarks of four different flavors.
The topology of recombination diagrams being different from that of
direct diagrams, the derivation of $T$-adequate sum rules
necessitates a more detailed study. The proof given here is based
on the analysis of four-quark singularites of Feynman diagrams.
This paper therefore completes the derivation of tetraquark-adequate
QCD sum rules.

The paper is organized as follows: Section \ref{1} highlights some
necessary properties of tetraquark interpolating currents and their
Green functions.
Section \ref{3} focuses on recombination Green functions, in which
quark flavors in the initial and final tetraquark currents
are arranged differently. Our conclusions follow in Section \ref{c}.

\section{\label{1}Tetraquark interpolating currents and correlation
functions}

We discuss properties of tetraquarks consisting of two quarks of
flavors $a$ and $c$ and two antiquarks of flavors $b$ and $d$.
The Dirac structure of the currents is of no relevance for the
arguments of this paper and will not be specified; we therefore do not
explicitly write the appropriate combinations of $\gamma$ matrices
between the quark fields.

We will exploit two properties of the exotic currents and their
Green functions:
\begin{itemize}
\item[(i)]
QCD sum rules adopt local multiquark interpolating
currents, and any local tetraquark current may be brought
to the form of a linear combination of products of color-singlet
combinations of quark fields with two different flavor
structures $\theta_{\bar a b\bar c d}=j_{\bar a b}j_{\bar c d}$ and
$\theta_{\bar a d\bar c b}=j_{\bar a d}j_{\bar c b}$ with $j_{ab}=\bar q_a q_b$ \cite{jaffe}.\footnote{
More generally, any gauge-invariant multiquark operator can be
reduced to a combination of products of colorless clusters
\cite{hagop2019}.}
For instance, a triplet-antitriplet tetraquark current may be written as
\begin{eqnarray}
(\epsilon^{ijk}\bar q_a^j\bar q_c^k)(\epsilon^{ij'k'}q_b^{j'}q_d^{k'})
=-\theta_{\bar a b\bar c d}-\theta_{\bar a d\bar c b}, 
\end{eqnarray}
and for an octet-octet current one finds   
\begin{eqnarray}
(\bar q_a T^A q_b)(\bar q_c T^A q_d)=
-\frac{1}{2}  \theta_{\bar a d\bar c b}- \frac{1}{6}  \theta_{\bar a b\bar c d}. %\nonumber
\end{eqnarray}
Here,  
$T^A=\lambda^A/2$; $A=1,\dots 8$, $\lambda^A$ are the Gell-Mann matrices, and
we made use of the anticommutativity of quark fields. 
Taking into account the Dirac structure of quark bilinears, one 
needs to perform also the Fierzing with respect to the spinor indices (see e.g. Refs.~\cite{Esposito:2016noz,QCD_SR_1}). 
 
It is therefore sufficient to study QCD sum rules for exotic
interpolating currents taken as products of two color-singlet
quark bilinears. (When needed, the momentum of a four-quark
current will be designated by $p$.)
\item[(ii)]
Any diagram involving the tetraquark currents
$\theta_{\bar a b\bar c d}$ and/or $\theta_{\bar a d\bar c b}$ can
be obtained from a diagram involving only the bilinear quark currents
$j_{\bar a b}$, $j_{\bar c d}$, $j_{\bar a d}$, and $j_{\bar c b}$
by merging the appropriate vertices. In particular, the two-point
function of tetraquark currents,
\begin{equation} 
\label{2e1}
\Pi_{\theta\theta}=
\langle {\rm T}\left\{\theta(x)\theta^{\dagger}(0)\right\}\rangle,
\end{equation}
can be obtained from the four-point function of ordinary bilinear currents
\begin{equation} \label{2e2}
\Gamma_{4j}=\langle {\rm T}\left\{j(x_1)j(x_2)
j^{\dagger}(x_3)j^{\dagger}(0)\right\}
\rangle
\end{equation}
by merging two pairs of vertices. The three-point function
involving one tetraquark current and two bilinear interpolating
currents,
\begin{equation} \label{2e3}
\Gamma_{\theta jj}=
\langle {\rm T}\left\{\theta(0)j^{\dagger}(x)j^{\dagger}(y)\right\}
\rangle,
\end{equation}
can be obtained from the same four-point function $\Gamma_{4j}$ by
merging only one pair of vertices. The functions  $\Pi_{\theta\theta}$ and
$\Gamma_{\theta jj}$ have been the subject of QCD sum rule analyses,
extensively presented in the literature.
\end{itemize}

According to property (ii), the quark-hadron duality relations for
$\Pi_{\theta\theta}$ and $\Gamma_{\theta jj}$ and the corresponding
QCD sum rules follow directly from the duality relations for
$\Gamma_{4j}$. Analytic properties of the latter, in particular, the
structure of its four-quark singularities, relevant for the derivation
of consistent QCD sum rules for tetraquarks, have been studied in
detail in \cite{lms_prd,lms_epjc}.

For a given global flavor content of the tetraquark current
$\bar a\bar c bd$, we have at our disposal two different flavor combinations of
two color singlets, $\theta_{\bar a b\bar c d}$ and
$\theta_{\bar a d\bar c b}$, and therefore one should distinguish
between the diagrams where quark flavors in the initial and
final states are combined in the same way (direct diagrams) and
in a different way (quark-exchange or recombination diagrams).
The Feynman diagrams for the corresponding four-point functions
have different topologies and structures in their four-quark
singularities and therefore necessitate separate studies.
For the case of direct diagrams, the reader is referred to Ref.
\cite{lms_prd100}, where a detailed analysis has been presented.
In the following, we concentrate on the case of quark-exchange
diagrams.

\section{Recombination Green functions involving tetraquark currents
\label{3}}
We now discuss diagrams with a recombination or quark-exchange
topology, where the quark flavors in the inital and the final
currents are arranged differently.

For the direct Green functions
\begin{equation} \label{3e0}
\Gamma_{4j}^{\rm dir}=  
\langle {\rm T}\left\{j_{\bar a b}(x_1)j_{\bar c d}(x_2)
j_{\bar a b}^{\dagger}(x_3)j_{\bar c d}^{\dagger}(0)\right\}
\rangle, 
\end{equation}
due to the factorization property of
lowest-order QCD diagrams, use was made of conventional SVZ sum rules
for correlators of bilinear quark currents to show the cancellation of the
non-T-phile diagrams on the OPE side against factorizable two-meson
contributions on the hadron side \cite{lms_prd100}. This transparent
procedure is not applicable in the recombination case and therefore
one needs to follow a different line of argument, based on the
singularity analysis of Feynman diagrams, for the related study.

Let us focus on the quark-hadron duality relations for the
recombination four-point function 
\begin{equation} \label{3e1}
\Gamma_{4j}^{\rm rec}=  
\langle {\rm T}\left\{j_{\bar a b}(x_1)j_{\bar c d}(x_2)
j_{\bar a d}^{\dagger}(x_3)j_{\bar c b}^{\dagger}(0)\right\}
\rangle.
\end{equation}
As emphasized in Sec.~\ref{1}, the understanding of the duality
relations for $\Gamma_{4j}^{\rm rec}$ immediately leads to
the understanding of the duality relations for
$\Pi_{\theta\theta}$ and $\Gamma_{\theta jj}$, since the diagrams
for the latter are obtained from the diagrams for
$\Gamma_{4j}^{\rm rec}$ by merging the appropriate vertices.

The analytic properties of the four-point functions of bilinear currents 
have been studied in detail in \cite{lms_prd,lms_epjc}. Here, we would like to 
show one example which is particularly important for the understanding of quark-hadron duality relations for the 
recombination diagrams. 

\begin{figure}
% Fig 1
\centerline{\includegraphics[width=7.cm]{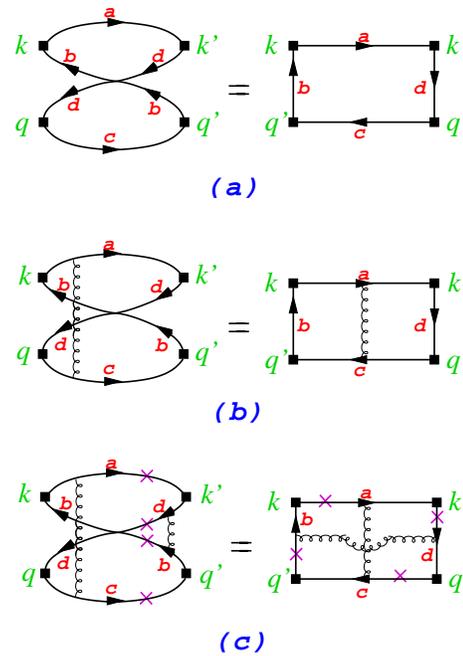}}
\caption{\label{Fig:4pt_s-cut}
Singularities of Feynman diagrams describing $\Gamma_{4j}^{\rm rec}$. 
The left-column (right-column) diagrams (a) and (b) do not contain a four-particle $s$-channel ($u$-channel) cut. 
The lowest-order diagram that has the four-particle $s$-channel cut is the left-column diagram (c). This cut [and the corresponding 
$u$-channel cut in the diagram (c) in the right column] emerges when the crossed quark propagators go on the mass shell.}
\end{figure}

\begin{figure*}[!t]
\includegraphics[width=12cm]{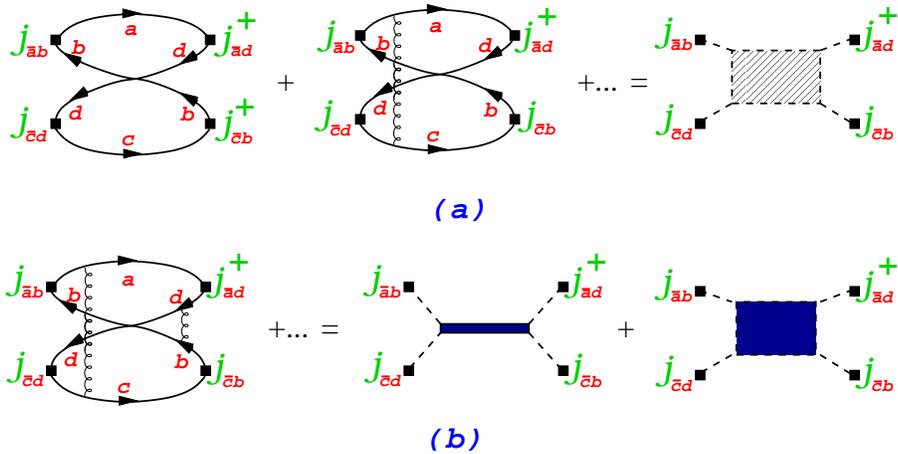}
\caption{\label{Fig:4pt_rec}
Duality relations for two classes of contributions to
$\Gamma_{4j}^{\rm rec}$:
(a) OPE representation for those contributions to
$\Gamma_{4j}^{\rm rec}$ that do not contain four-quark $s$-channel cuts
(non-$T$-phile contributions) and the corresponding representation for
such contributions using hadron degrees of freedom: the dashed
rectangle denotes the sum of all contributions to
$\Gamma_{4j}^{\rm rec}$ that do not have two-meson $s$-channel cuts.
(b) OPE representation for those contributions to
$\Gamma_{4j}^{\rm rec}$ that do have four-quark $s$-channel cuts
(class-$T$ contributions) and the corresponding representation for
class-$T$ contributions using meson degrees of freedom: the solid
blue rectangle denotes the sum of all hadronic contributions to
$\Gamma_{4j}^{\rm rec}$ that have two-meson $s$-channel cuts.
A possible isolated $T$-pole has been added.}
\end{figure*}

Figure \ref{Fig:4pt_s-cut} presents the recombination diagrams of lowest 
orders in an unfolded form as box diagrams:  
the $s$-channel singularities of the diagrams in the left column correspond to $u$-channel 
singularities of the diagrams 
in the right column. 
The absence of $u$-channel cuts (not only four-quark cuts, but any cuts) 
in the diagrams in the right column 
of Fig.~\ref{Fig:4pt_s-cut}(a,b) is evident; this means the absence of $s$-channel 
singularities in the diagrams 
Fig.~\ref{Fig:4pt_s-cut}(a,b) in the left column. 
In order to understand the structure of singularities of the diagram with two-gluon exchanges, 
Fig.~\ref{Fig:4pt_s-cut}(c), 
one needs to solve the Landau equations \cite{landau}. 
The corresponding equations and their solutions are presented in the Appendix. 
With the help of the Landau equations one finds that the left-column diagram Fig.~\ref{Fig:4pt_s-cut}(c) 
has the four-quark $s$-channel threshold at $s=(m_a+m_b+m_c+m_d)^2$; this threshold and the
corresponding four-quark cut is related 
to the configuration of quark momenta when the crossed quark lines go on their mass shell. 
With this knowledge at hand, we turn to the analysis of QCD sum rules for the recombination Green functions. 
Figure~\ref{Fig:4pt_rec} presents in diagrammatic form the quark-hadron duality relations for $\Gamma_{4j}^{\rm rec}$.

% Figs 8

Several remarks are in order.
\begin{itemize}
\item[(i)]
All QCD diagrams for $\Gamma_{4j}^{\rm rec}$ may be divided into two
non-overlapping classes according to the structure of their $s$-channel
singularities: The diagrams of the first class, referred to as
non-$T$-phile diagrams, do not have a four-quark $s$-channel cut.
All diagrams that contain a four-quark $s$-channel cut belong to the
second class, referred to as $T$-phile diagrams. If a tetraquark
pole emerges in the $s$-channel, then it can emerge only through
the infinite set of $T$-phile diagrams. Any $s$-channel singularity
in the set of non-$T$-phile diagrams is not related to the tetraquark.
\item[(ii)]
If we look at the representation of Green functions in the hadron
picture, then Green functions of the non-$T$-phile class do not
contain two-meson $s$-channel intermediate states, whereas the hadron
representation for $T$-phile contributions does contain such
two-meson intermediate states and also a possible tetraquark state.
\item[(iii)]
Quark-hadron duality relations are fulfilled for non-$T$-phile
diagrams and for $T$-phile diagrams separately.
It is therefore straightforward to write down the corresponding QCD
sum rules, as in Fig.~\ref{Fig:4pt_rec}.
\end{itemize}

In Fig.~\ref{Fig:4pt_rec}(a), the non-$T$-phile diagrams on the OPE
side are dual to the specific meson diagrams without two-meson
$s$-channel cuts and without a possible tetraquark pole on the hadron
side of the QCD sum rule for $\Gamma_{4j}^{\rm rec}$. (Two-meson
cuts appear in the $t$- and $u$-channels.) Obviously, non-$T$-phile
diagrams cannot be related to the tetraquark properties.

In Fig.~\ref{Fig:4pt_rec}(b), the QCD sum rule for the $T$-phile part
of $\Gamma_{4j}^{\rm rec}$, represents the desired $T$-adequate QCD
sum rule.

\begin{figure}[!b]
% Fig 9
\centerline{\includegraphics[width=8.5cm]{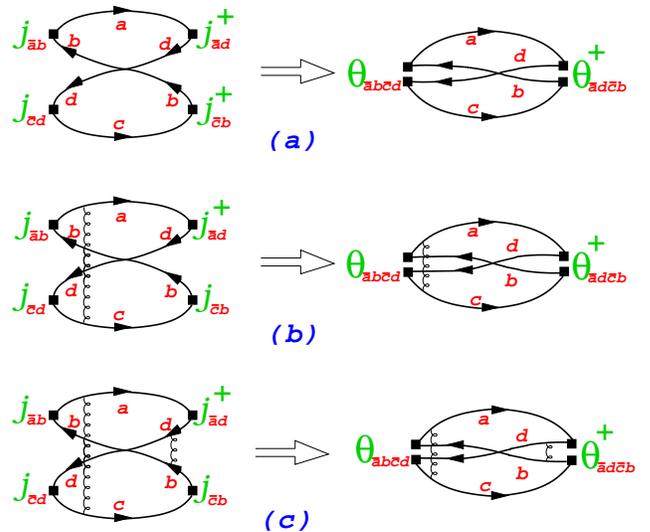}}
\caption{\label{Fig:2pt_rec}
Feynman diagrams for a recombination two-point function of tetraquark
currents (r.h.s.). They are obtained by merging vertices in the
recombination four-point function of bilinear quark currents (l.h.s.).
Diagram (c) on the l.h.s.\ is the lowest-order diagram that contains
a four-quark $s$-channel cut. Diagrams (a) and (b) on the l.h.s.\ do
not contain four-quark singularities in the $s$-channel. Only
diagram (c) on the l.h.s.\ is a $T$-phile diagram. Similarly, 
among the diagrams on the r.h.s., it is only diagram (c) that
contributes to the OPE side of the tetraquark SVZ sum rule (\ref{3.1}).}
\end{figure}

\begin{figure}%[!h]

% Fig 10
\centerline{\includegraphics[width=8cm]{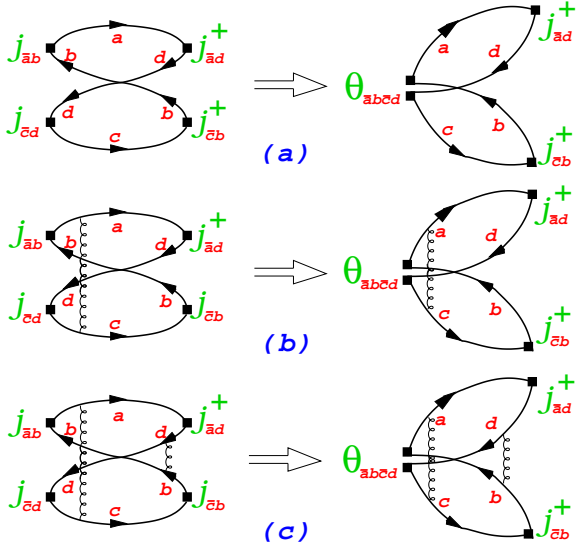}}
\caption{\label{Fig:3pt_rec}
Feynman diagrams for a recombination three-point function (r.h.s.).
They are obtained by merging vertices in the four-point function of
bilinear quark currents (l.h.s.). Diagram (a) in the r.h.s. has at
most a polynomial dependence on the variable $p^2$
($p$: the momentum of the four-quark current) and does not
have a four-quark cut in it. Diagram (b) in the r.h.s.\ has a
nontrivial $p^2$ dependence, but it drops out from the
$T$-adequate QCD sum rule: its contribution cancels against the
hadron contributions not related to the tetraquark properties by
virtue of the duality relations for $\Gamma_{4j}^{\rm rec}$.
Therefore, diagrams (a) and (b) are not related to tetraquark
properties. Only the diagram with the two-gluon exchange (c) is a
$T$-phile diagram, contributing to the tetraquark SVZ sum rule
(\ref{3.2}).}
\end{figure}

Having at hand the latter relation, one easily constructs $T$-adequate
QCD sum rules for $\Pi_{\theta\theta}^{\rm rec}$ and
$\Gamma_{\theta jj}^{\rm rec}$ by merging the appropriate vertices in
$\Gamma_{4j}^{\rm rec}$, as shown in Figs. \ref{Fig:2pt_rec} and
\ref{Fig:3pt_rec}.
In the end, only the $T$-phile diagrams with, at least, two gluon
exchanges of the type shown in Fig.~\ref{Fig:2pt_rec}(c) and
Fig.~\ref{Fig:3pt_rec}(c) appear in the $T$-adequate QCD sum rules
for the tetraquark properties.

%%%%%%%%%%%%%%%%%%%%%%%%%%%%%%%%%%%%%%%%%%%%%%%%%%%%%%%%%%%%%%
%
Here, one should bear in mind a subtlety related to the difference between
confined (compact) tetraquarks and molecular tetraquark states: 
the compact tetraquarks are poles in full QCD Green functions, but they do not emerge in 
the effective low-energy theory described in terms of meson degrees of freedom; 
molecular states, on the contrary, are not only poles in full QCD Green functions, but also emerge as 
poles in the effective meson low-energy theory. Therefore, in the case of a molecular tetraquark state, 
the corresponding pole is contained in the infinite sum of diagrams denoted by the solid blue
rectangle of Fig.~\ref{Fig:4pt_rec}(b). The decomposition in the
r.h.s.\ of Fig.~\ref{Fig:4pt_rec}(b) should then be understood
as the sum of meson diagrams, from which the $T$-pole has already been subtracted. 
In case the tetraquark pole emerges as a compact four-quark state, it is not present in the 
effective meson theory and should be added separately, as in Fig.~\ref{Fig:4pt_rec}(b).

\begin{figure}
% Fig 11
\centerline{\includegraphics[width=8cm]{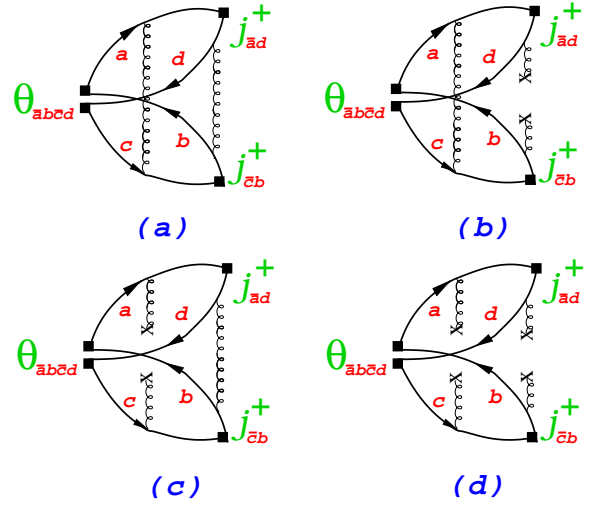}}
\caption{\label{Fig:3pt_rec_T}
Typical $T$-phile $O(\alpha_s^2)$ perturbative diagram (a) and the
corresponding power corrections of order
$O\left(\alpha_s\langle \alpha_sGG\rangle\right)$ (b,c) which
contribute to the OPE side of the $T$-adequate sum rule for the
$T\to M_{\bar a d}M_{\bar c b}$ coupling. Diagram (d), of order
$O(\langle\alpha_s^2GGGG\rangle)$, has at most a polynomial dependence
on $p^2$ and thus does not contribute to the Borel sum rule for
the $T\to M_{\bar a d}M_{\bar c b}$ coupling.}
\end{figure}
As a final step, assuming that the high-momentum tail (corresponding
to the $s$-integration above an effective threshold $s_{\rm eff}$ \cite{lms_seff1,lms_seff2,lms_seff3}) of
the $T$-phile Feynman diagrams on the OPE side of QCD sum rules cancels
against the hadron continuum contributions on the hadron side
of QCD sum rules, we arrive at the ultimate $T$-adequate QCD sum rules
for recombination Green functions
%\newpage
\begin{eqnarray}
\label{3.1}
&&f^{\bar ab\bar cd}_Tf^{\bar ad\bar cb}_T\exp(-M_T^2\tau)\nonumber\\
&&\ \ \ =\int\limits_{(4m_q)^2}^{s_{\rm eff}} ds
\exp(-s \tau)\rho^{\rm rec}_{T}(s)+\mathrm{BPC},
\end{eqnarray}
\begin{eqnarray}
\label{3.2}
&&f^{\bar ab\bar cd}_T A(T\to j_{\bar ad}j_{\bar cb})
\exp(-M_T^2\tau)\nonumber\\
&&\ \ \ =\int\limits_{(4m_q)^2}^{s_{\rm eff}} ds
\exp(-s \tau)\Delta^{\rm rec}_{T}(s)+\mathrm{BPC},
\end{eqnarray}
\begin{equation}
\label{3.3}
f^{\bar ab\bar cd}_T=\langle T|\theta_{\bar ab\bar cd}|0\rangle,
\qquad f^{\bar ad\bar cb}_T=\langle T|\theta_{\bar ad\bar cb}|0\rangle,
\end{equation}
where BPC is short for Borelized power corrections, 
$4m_q\equiv m_a+m_b+m_c+m_d$, and
$A(T\to j_{\bar ad}j_{\bar cb})$ is the amplitude
$\langle 0|{\rm T}\{ j_{\bar ad}(x) j_{\bar cb}(0)\}|T(p)\rangle$
in momentum space; $\rho^{\rm rec}_{T}(s)$ and
$\Delta^{\rm rec}_{T}(s)$ are the
spectral densities in the variable $s$ of the $O(\alpha_s^2)$ diagrams
with two-gluon exchanges, of the type shown in the r.h.s. 
of Fig.~\ref{Fig:2pt_rec}(c) and Fig.~\ref{Fig:3pt_rec}(c),
respectively.
Power corrections in the r.h.s.\ of Eqs.~(\ref{3.1}) and (\ref{3.2})
correspond to condensate insertions in these diagrams. The typical
lowest-order diagrams contributing to the OPE side of
Eq.~(\ref{3.2}) are shown in Fig.~\ref{Fig:3pt_rec_T}(a,b,c);
the power correction given by Fig.~\ref{Fig:3pt_rec_T}(d)
does not depend on $p^2$ and thus vanishes under the Borel transformation 
and does not contribute to the Borel sum rule (\ref{3.2}).

\begin{figure}[!b]
\centerline{\includegraphics[width=8.5cm]{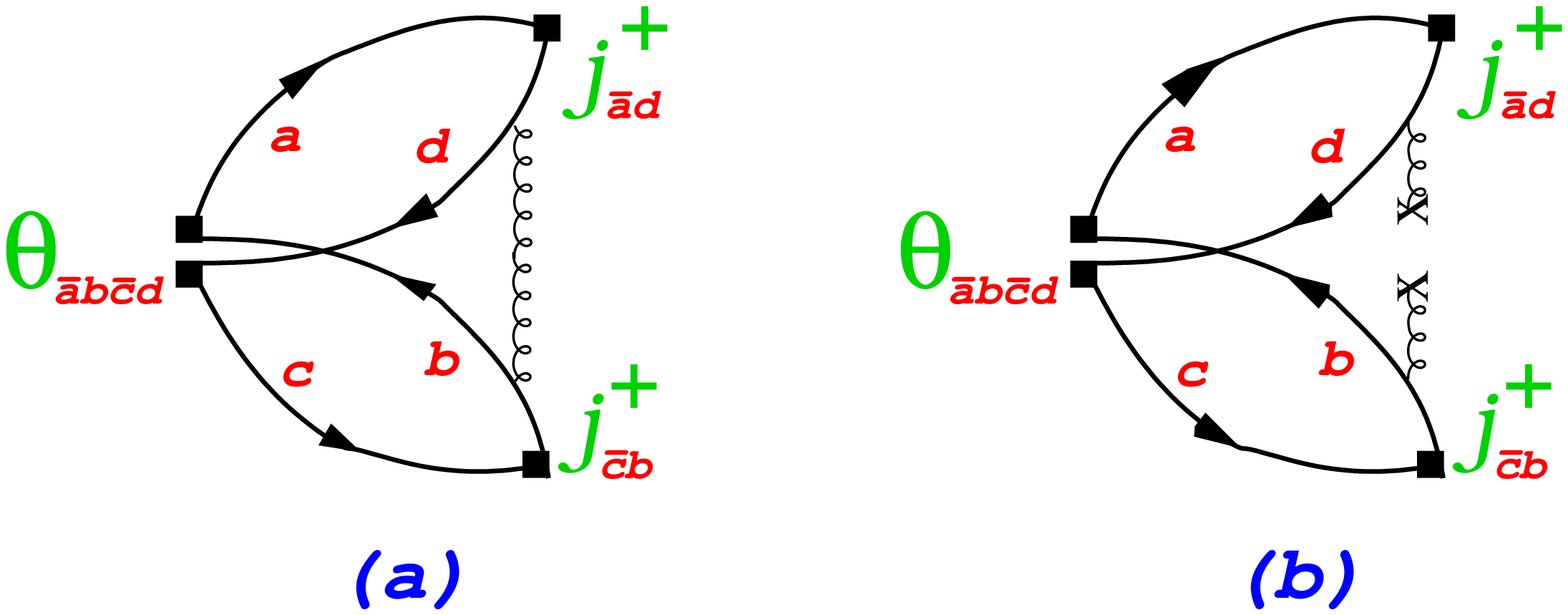}}
\caption{\label{Fig:3pt_rec_non}
A typical perturbative $O(\alpha_s)$ diagram and
$O\left(\langle \alpha_sGG\rangle\right)$ power correction that
both do not contribute to the OPE side of the $T$-adequate QCD sum
rule for the $T\to M_{\bar a d}M_{\bar c b}$ coupling, as these
diagrams are obtained by merging vertices in the non-$T$-phile diagram
for $\Gamma^{\rm rec}_{4j}$.}
\end{figure}
%%%%%%%%%%%%%%%%%%%%%%%%%%%%%%%%%%%%%%%%%%%%%%%%%%%%%%%%%%%%%%%%%%%%%%
%%%%%%%%%%
Typical diagrams that do not contribute to the OPE side of the
$T$-adequate QCD sum rules for $\Gamma_{\theta jj}^{\rm rec}$
are shown in Fig.~\ref{Fig:3pt_rec_non}. These have often been
considered in the literature as the appropriate OPE diagrams for
the coupling of a tetraquark to two mesons,
$T\to M_{\bar a d}M_{\bar c b}$ \cite{QCD_SR_1}.
However, as it is clear from the duality relation depicted in 
Fig.~\ref{Fig:4pt_rec}(a), in spite of a nontrivial $p^2$ dependence
of the diagram in Fig.~\ref{Fig:3pt_rec_non}(a), it does not
contribute to the $T$-adequate QCD sum rule for
$\Gamma_{\theta jj}^{\rm rec}$. For the same reason, the condensate
diagram Fig.~\ref{Fig:3pt_rec_non}(b) also cannot contribute to the
$T\to M_{\bar a d}M_{\bar c b}$ coupling. (Moreover, the diagram
Fig.~\ref{Fig:3pt_rec_non}(b) does not depend on $p^2$ at all and
thus its Borel transform vanishes.)
%%%%%%%%%%%%%%%%%%%%%%%%%%%%%%%%%%%%%%%%%%%%%%%%%%%%%%%%%%%%%%%%%%

We have discussed the case of flavor-exotic tetraquark currents.
It is clear, however, that the same arguments apply also to
the case of crypto-exotic flavor structure of the tetraquark current
(i.e., $\bar q_a q_b \bar q_b q_c$, with $a,b,c$ denoting quark
flavors). Compared to the flavor-exotic case, the crypto-exotic case
exhibits an extended set of $T$-phile diagrams (see a detailed discussion
in \cite{lms_epjc}). Apart from this feature, the splitting of all
diagrams in the OPE for $\Gamma_{4j}$ into two non-overlapping classes
of non-$T$-phile and $T$-phile diagrams is valid also in the
crypto-exotic case. Obviously, the arguments given in this section
remain unchanged; the only modification appears in the way of
selecting the appropriate set of $T$-phile diagrams.

Considering tetraquark interpolating currents in the form of 
products of two color-singlet bilinears allowed us to formulate
a clear criterion for selecting $T$-phile diagrams that contribute
to the tetraquark-adequate QCD sum rules for $\Pi_{\theta\theta}$ and
$\Gamma_{\theta jj}$. One may use other color structures of the quark
bilinears forming colorless local tetraquark interpolating currents:
for instance, one may work with the triplet-antitriplet,
i.e., diquark-antidiquark, structure $\bar D^i D_i$ with
$D_i=\epsilon_{ijk} q^j q^k$, $i,j,k=1,2,3$ being color indices,
or the octet-octet structure
$\bar q \lambda^A q \bar q \lambda^A q$, $\lambda^A$ being the
Gell-Mann matrices, $A=1,\dots,8$.
However, for such color structures of the tetraquark interpolating
currents, it is difficult to provide explicit criteria for selecting
the appropriate $T$-phile diagrams. A consistent way to select such
$T$-phile diagrams is to rearrange the diquark-antidiquark or
octet-octet local currents into the singlet-singlet color structures
and make use of the criteria for selecting the $T$-phile diagrams
already formulated for this case. In this way, the derivation
of the tetraquark-adequate QCD sum rules also works for other
choices of the color structure of the local tetraquark interpolating
currents.

%***************************************************************
\section{Conclusion}
\label{c}

We have considered the SVZ sum rules for correlation functions
involving exotic tetraquark currents $\theta$, namely, the two-point
function $\Pi_{\theta\theta}$, and the three-point function
$\Gamma_{\theta jj}$, specifically related in the present paper
to quark-exchange processes.

It turns out that the duality relations and QCD sum rules for
these correlators have different properties compared with the
duality relations for the correlators of bilinear quark currents.
The OPE part as well as the hadron part of the sum rules split
into two non-overlapping classes of contributions; each of them
satisfies its own QCD sum rule:
One QCD sum rule contains, on the OPE side, tetraquark-concerning
diagrams, also named $T$-phile diagrams, which start at order
$O(\alpha_s^2)$, and on the hadron side, the tetraquark contribution.
This sum rule has been named $T$-adequate sum rule.
The second QCD sum rule, which contains non-$T$-phile diagrams on
the OPE side, does not contain the tetraquark contribution on the
hadron side and therefore has no relation to the tetraquark properties.

The $T$-phile diagrams for any correlator involving the tetraquark
currents can be obtained from the corresponding $T$-phile diagrams
of the four-point function $\Gamma_{4j}$ of quark color-singlet
bilinear currents; the latter are defined as those diagrams which
have four-quark $s$-channel cuts. The $T$-phile diagrams can be
identified by solving the Landau equations \cite{landau}.

The present work completes the proof provided in \cite{lms_prd100}
for direct-type processes, where the quark flavors are arranged in
color-singlet bilinears in the same way in the initial and in the
final states.

\begin{acknowledgments}
The authors are grateful to T.~Cohen, F.-K.~Guo, M.~Knecht, L.~Maiani,
B.~Moussallam, A.~Polosa, V.~Riquer, S.~Simula, B.~Stech, and W. Wang
for valuable discussions. D.~M.~acknowledges support from the Austrian
Science Fund (FWF), project~P29028. H.~S.~acknowledges support from
the EU research and innovation programme Horizon 2020, under grant
agreement No 824093. D.~M. and H.~S. are grateful for support under
joint CNRS/RFBR grant PRC Russia/19-52-15022.
\end{acknowledgments}

%****************************************
\appendix
\section{Landau equations for recombination diagrams \label{appendix}}
A generic expression of a Feynman diagram has the form 
\begin{eqnarray} 
\label{ea1}
I(p)=\int\prod_{\ell=1}^{L}\frac{d^4k_{\ell}^{}}{(2\pi)^4}
\prod_{i=1}^{I}\frac{1}{(q_i^2-m_i^2+i\epsilon)},
\end{eqnarray}
where $p$ represents a set of external momenta and $q_i$ are linear 
functions of the $p$'s and of the independent loop variables $k$.

The Landau equations are \cite{landau}
\begin{eqnarray}
\label{ea2}
\lambda_i^{}(q_i^2-m_i^2)=0,\qquad i=1,\ldots,I,\\
\label{ea3}
\sum_{i=1}^I\lambda_i q_i^{}\cdot\frac{\partial q_i^{}}
{\partial k_{\ell}^{}}=0,\qquad \ell=1,\ldots,L,
\end{eqnarray}  
where the $\lambda$'s are parameters (Lagrange multipliers) to be
determined.

This system of equations may have independent subsystems, corresponding
to the vanishing of a certain number of parameters $\lambda$.

Here, we present the Landau equations for the Feynman diagrams shown in the left column of 
Fig.~\ref{Fig:4pt_s-cut}.
We are interested in the singularities produced by the quark propagators. We therefore do not 
consider gluon propagators in the Landau equations;  
this amounts to taking the corresponding $\lambda$'s equal to zero.

We start with the recombination diagram of leading order of Fig.~7(a).%\ref{Fig:4pt_s-cut_landau}(a). 
\begin{figure}
\label{Fig:4pt_s-cut_landau}
\centerline{\includegraphics[height=14cm]{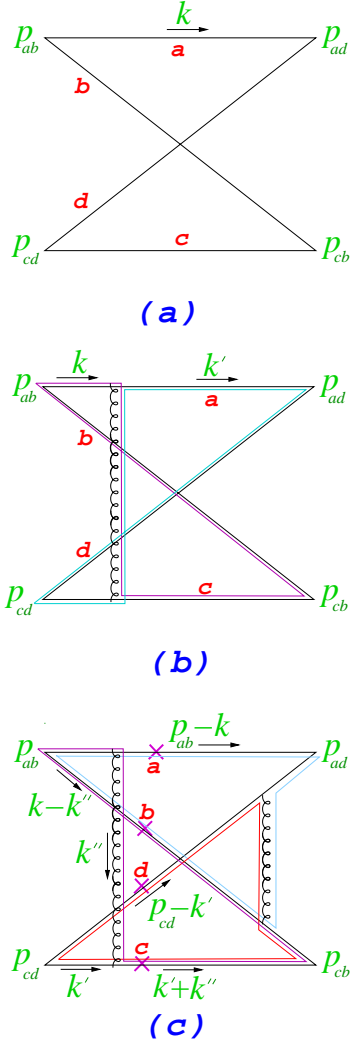}}
\caption{Diagrams of Fig.~\ref{Fig:4pt_s-cut} with explicit momentum flow.}
\end{figure}
The external momenta are $p_{ab}^{}$, $p_{cd}^{}$, $p_{ad}^{}$, $p_{cb}^{}$, with 
\begin{eqnarray}
\label{ea4}
& &P=p_{ab}^{}+p_{cd}^{}=p_{ad}^{}+p_{cb}^{},\ \ \ \ \ 
s=P^2,\nonumber \\
& & t=(p_{ab}^{}-p_{ad}^{})^2,\ \ \ \ \ \ u=(p_{ab}^{}-p_{cb}^{})^2.
\end{eqnarray} 
The Landau equations read
\begin{eqnarray}
\label{ea5}
&&\lambda_a^{}(k^2-m_a^2)=0, \nonumber\\
&&\lambda_b^{}((k-p_{ab})^2-m_b^2)=0,\nonumber \\
&&\lambda_c^{}((p_{cd}^{}-p_{ad}^{}+k)^2-m_c^2)=0, \nonumber\\
&&\lambda_d^{}((k-p_{ad}^{})^2-m_d^2)=0,\nonumber \\ 
&&\lambda_a^{}k+\lambda_b^{}(k-p_{ab})\nonumber\\
&&\qquad +\lambda_c^{} (k+p_{cd}^{}-p_{ad})+\lambda_d^{}(k-p_{ad}^{})=0.
\end{eqnarray}
This system of equations has several independent subsystems.
Choosing $\lambda_b^{}=\lambda_d^{}=0$, one obtains 
$u=(m_a^{}\pm m_c^{})^2$.
(Only physical singularities, corresponding to $+$ signs between the masses, are relevant.)
Choosing $\lambda_a^{}=\lambda_c^{}=0$, one obtains 
$t=(m_b^{}\pm m_d^{})^2$.
Choosing $\lambda_c^{}=\lambda_d^{}=0$, one obtains 
$p_{ab}^2=(m_a\pm m_b)^2$, and so forth. 
%The latter type of singularities are present inside the meson propagators. 
The property that the singularities in $u$ and $t$ involve only two quark masses shows that there 
are no four-quark singularities in this diagram. In the variable $s$ no singularities at all are found.

%**********************************
The second diagram, Fig.\ 7(b), is a two-loop (the loops shown in different colors) 
diagram with one-gluon exchange between quarks $a$ and $c$. 
It contains seven propagators and thus the full system of Landau equations includes 
seven parameters $\lambda_i$. 
One has two four-momentum conservation laws related to two loops. The full system of Landau 
equations splits into several subsystems 
related to setting to zero some of the parameters $\lambda_i$. 
These subsystems lead to thresholds at 
$p_{ij}^2=(m_i\pm m_j)^2$, $t=(m_b\pm m_d)^2$ and $u=(m_a\pm m_c)^2$ 
(Only physical singularities related to $+$ signs between the masses are relevant).  
None of the subsystems leads to the solution corresponding to the threshold in the variable $s$, 
indicating the absence of $s$ cuts. 
%**********************************
Finally, we turn to the diagram Fig.~7(c) with two-gluon exchanges between quarks 
$a$ and $c$, and $b$ and $d$. It is a three-loop diagram (each loop shown in a different color), 
containing three independent integration 
momenta $k$, $k'$, and $k''$, and ten propagators.
Therefore, the general system of Landau equations contains ten parameters $\lambda_i$ (equal to 
the number of the propagators) and three independent 
momentum conservation relations, equal to the number of loops. 
We present a subsystem of Landau equations corresponding to the 
crossed propagators in Fig.\ 7(c) (i.e., all other $\lambda$'s are set to zero) 
which leads to the four-quark threshold in the variable $s$: 
\begin{eqnarray}
\label{ea11}
& &\lambda_a^{}((p_{ab}^{}-k)^2-m_a^2)=0,\ 
\lambda_b^{}((k-k'')^2-m_b^2)=0,\nonumber \\
& &\lambda_c^{}((k'+k'')^2-m_c^2)=0,\ 
\lambda_d^{}((p_{cd}-k')^2-m_d^2)=0,\nonumber \\
& & \\ 
\label{ea12}
& &\lambda_a^{}(p_{ab}^{}-k)-\lambda_b^{}(k-k'')=0,\nonumber \\ 
& &\lambda_c^{}(k'+k'')-\lambda_d^{}(p_{cd}-k')=0,\nonumber \\ 
& &-\lambda_b^{}(k-k'')+\lambda_c^{}(k'+k'')=0.
\end{eqnarray}
This system of four equations can be solved and leads to a nontrivial solution 
$P^2=s=(m_a^{}+m_b^{}+m_c^{}+m_d^{})^2$, thus indicating the presence of an $s$-channel 
four-quark threshold. (The unphysical
singularities, corresponding to changes of sign in front of the 
masses, exist as well.) 

%\input{Appendix.tex}
% Create the reference section using BibTeX:
%\bibliography{basename of .bib file}

%\input{Appendix.tex}
\end{document}